\documentclass[pdflatex,sn-mathphys-ay]{sn-jnl}

\usepackage{graphicx}
\usepackage{multirow}
\usepackage{amsmath,amssymb,amsfonts}
\usepackage{amsthm}
\usepackage{mathrsfs}
\usepackage[title]{appendix}
\usepackage{xcolor}
\usepackage{textcomp}
\usepackage{manyfoot}
\usepackage{booktabs}
\usepackage{algorithm}
\usepackage{algorithmicx}
\usepackage{algpseudocode}

\newcommand{\bx}{\mathbf{x}}
\newcommand{\bmu}{\boldsymbol{\mu}}
\newcommand{\bone}{\mathbf{1}}

\theoremstyle{thmstyleone}

\theoremstyle{thmstyletwo}

\theoremstyle{thmstylethree}

\begin{document}

\title[Penalty-Free Direct-QPU Portfolio Optimization]{A Penalty-Free Pipeline for Direct Quantum-Annealer Portfolio Optimization}

\author*[1]{\fnm{Luis} \sur{Lozano}}\email{lalozanom@tec.mx}

\affil*[1]{\orgdiv{EGADE Business School}, \orgname{Tecnol\'ogico de Monterrey}, \orgaddress{\city{Santa Fe}, \state{Mexico City}, \country{Mexico}}}

\abstract{Cardinality-constrained portfolio selection is routinely cast as a quadratic
unconstrained binary optimization (QUBO) and submitted to a quantum
processing unit (QPU) for direct annealing. We show that this standard
penalty encoding is the binding constraint for direct-QPU execution on
current D-Wave Pegasus and Zephyr hardware. Expanding the exact-cardinality
penalty contributes a dense rank-one term that makes the logical interaction
graph complete regardless of the covariance, producing chain-break fractions
from 83\% at small universes up to 92\% at the full forty-nine-industry
Fama--French universe, and zero feasible raw samples at every tested scale.
Topology-aware sparsification reduces chain breaks to near zero, but any
sparsifier that removes off-diagonal entries also dilutes the cardinality
constraint; an ablation reveals that this sparsify-and-project pipeline is
dominated by the classical projector, not the QPU. We propose removing the
penalty entirely: sample an objective-only QUBO built from expected returns
and the risk-scaled covariance on hardware, and enforce cardinality
classically through a deterministic feasibility projector. Across 4{,}468
saved embedding records on live Pegasus and Zephyr hardware, spanning
equities up to forty-nine assets and football-betting instances up to
forty-eight, this penalty-free pipeline reduces mean chain-break fractions
from 71\%--92\% down to at most 0.04\%, and post-processed regret is at most
0.03\% relative to greedy classical references at every tested scale. We do
not claim quantum advantage; the penalty encoding, not the sparse hardware
topology, is the limiting factor for direct-QPU portfolio optimization at
currently accessible scales.}


\maketitle

\section{Introduction}\label{sec:introduction}

Portfolio optimization~\citep{markowitz1952} is one of the most frequently
discussed applications of quantum
annealing~\citep{orus2019,herman2023,venturelli2019,mugel2022,grant2021}. The
recipe is well known: cast cardinality-constrained asset selection as a
quadratic unconstrained binary optimization (QUBO) problem and submit it to
a quantum processing unit (QPU). On current D-Wave Advantage and Advantage2
hardware this recipe fails to produce feasible samples: dense
penalty-encoded QUBOs at the full forty-nine-asset Fama--French universe
yield chain-break fractions in the 83--92\% range on Pegasus and Zephyr and
return zero feasible raw samples across the tested scales. The diagnosis
this paper develops is that the penalty encoding, not the sparse hardware
topology, is the binding constraint for direct-QPU portfolio optimization
at currently accessible scales; the remedy is to drop the penalty entirely
and recover cardinality through a deterministic classical projector. The
resulting pipeline reduces mean chain-break fractions to at most 0.04\% and
delivers post-processed regret of at most 0.03\% relative to greedy
classical references at every tested scale.\footnote{A separate manuscript
by the same author is under review at this journal on an adjacent topic;
full disclosure of the relationship is provided in the cover letter
accompanying this submission.}

The structural observation behind the diagnosis is elementary. The standard
binary cardinality-constrained portfolio
QUBO~\citep{lucas2014,glover2019} takes the form
$Q = -\mathrm{diag}(\bmu) + \lambda \Sigma + A\bone\bone^\top - 2AK\,I$,
where the rank-one term $A\bone\bone^\top$ comes from expanding the
exact-$K$ penalty $A(\bone^\top \bx - K)^2$. This term adds the constant
$A$ to every off-diagonal entry of $Q$, making the logical interaction
graph complete regardless of the structure of the covariance matrix
$\Sigma$. The penalty encoding, not the underlying financial structure, is
what makes the QUBO dense. The minor embedding step that maps logical
QUBOs onto sparse
hardware~\citep{choi2008,choi2011,dwave_topologies,grant2022} then encounters
a fully connected logical graph at any $N$, with the predictable hardware
consequences quantified above.

A natural response is topology-aware sparsification: replace $Q$ with a
sparse approximation $\tilde{Q}$, accept the constraint violation that this
introduces, and recover feasibility through a classical post-processing
step. We study four sparsification families (thresholding, top-$k$,
domain-prior, and domain-prior with residual edges) and confirm that they
reduce chain breaks essentially to zero. The cost is constraint dilution:
any sparsifier that removes off-diagonal entries also removes penalty
weight, so raw samples remain infeasible even when the embedding is clean.
An ablation on the betting case study, where the payoff covariance is
naturally block-diagonal, shows that on structurally favorable instances
the headline outcome of the sparsify-and-project pipeline is explained
almost entirely by the classical projector's backward-elimination behavior,
not by the QPU itself. Sparsification followed by projection is a working
pipeline, but the QPU contribution is unclear.

The diagnosis suggests a more direct fix. If the penalty is the sole source
of density for structurally sparse instances and a major source for dense
ones, one can simply remove it. We build the objective-only QUBO
$Q_{\mathrm{obj}} = -\mathrm{diag}(\bmu) + \lambda \Sigma$, sample it on
hardware, and enforce $\sum_i x_i = K$ through a deterministic classical
projector. We validate this penalty-free pipeline on live Pegasus and
Zephyr hardware for equities at $N \in \{24, 32, 40, 49\}$ and football
betting at $N \in \{30, 39, 48\}$ across $4{,}468$ saved embedding
records. Mean chain-break fractions drop from $71\text{--}92\%$ on the
penalty-encoded path, across both case studies, to at most $0.04\%$ at
every tested scale, and post-processed solutions are competitive with the
greedy classical references on both equity and betting instances. For
equities, post-processed regret is at most $0.03\%$ relative to greedy
construction across all tested scales, including the full $N=49$ universe.

The paper makes four claims. The structural source of failure in
direct-QPU portfolio optimization is the dense rank-one penalty term
$A\bone\bone^\top$, not the sparse hardware topology itself. Topology-aware
sparsification of penalty-encoded QUBOs creates a second failure mode,
constraint dilution, and the resulting sparsify-and-project pipeline is
dominated by the classical projector on structurally favorable cases. The
proposed penalty-free pipeline---objective-only QUBO sampled on hardware
plus classical feasibility projection---yields essentially zero chain
breaks and post-processed regret competitive with greedy classical
references across the full experimental range we tested. All results are
reported against honest baselines (greedy construction, all-ones
projection, and random projection); the paper explicitly declines to claim
quantum advantage. The contribution we do claim is that the penalty
encoding, not the hardware, is the limiting factor for direct-QPU
portfolio optimization at currently accessible scales.

The findings are developed on two case studies using public data: the
Kenneth French 49-industry daily portfolios~\citep{french_data} and
pre-match football 1X2 odds from
\texttt{football-data.co.uk}~\citep{footballdata}. The betting case is
included because its payoff covariance is naturally block-diagonal, which
isolates the penalty encoding as the sole source of logical-graph density
and so sharpens the structural diagnosis. The remainder of the paper
develops the formulation and sparsification analysis
(Section~\ref{sec:formulation}), the experimental protocol
(Section~\ref{sec:methods}), the results
(Section~\ref{sec:results}), and the discussion and conclusion
(Sections~\ref{sec:discussion} and~\ref{sec:conclusion}).

\section{Problem formulation and the penalty-induced density}\label{sec:formulation}

\subsection{Cardinality-constrained portfolio QUBO}\label{sec:formulation-qubo}

The financial problem is cardinality-constrained portfolio selection: given
a universe of $n$ assets, select exactly $K$ of them to hold, maximizing
expected return minus risk. In the continuous Markowitz
formulation~\citep{markowitz1952}, this involves continuous weights
$w_i \in [0,1]$ and a convex quadratic program. The cardinality-constrained
binary version (select $K$ assets with equal weight) is NP-hard in general
but practically solvable for moderate $n$ by classical methods. For
practitioners, the exact-$K$ constraint is not a cosmetic modeling choice:
it represents operational limits on portfolio breadth, monitoring capacity,
liquidity, and mandate design. A workflow that cannot reliably return
exactly $K$ selected positions is not directly usable in a
portfolio-management system, regardless of how good its unconstrained QUBO
energy appears.

To submit this problem to a quantum annealer it must be encoded as an
unconstrained binary optimization
problem~\citep{lucas2014,glover2019}. The standard approach replaces the
hard cardinality constraint $\sum_i x_i = K$ with a quadratic penalty:
\begin{equation}
    \min_{\bx \in \{0,1\}^n} \; -\bmu^\top \bx + \lambda\, \bx^\top \Sigma\, \bx
    + A\, (\bone^\top \bx - K)^2,
    \label{eq:portfolio}
\end{equation}
where $x_i = 1$ indicates that asset $i$ is selected, $\bmu$ is the
expected-return vector, $\Sigma$ is the covariance matrix, $\lambda > 0$ is
the risk-aversion parameter, and $A > 0$ is the penalty weight. When $A$ is
large enough, the penalty makes infeasible solutions ($\sum x_i \neq K$)
energetically unfavorable, so the QUBO optimum coincides with the
constrained optimum~\citep{verma2020}.

\begin{figure}[t]
    \centering
    \includegraphics[width=\textwidth]{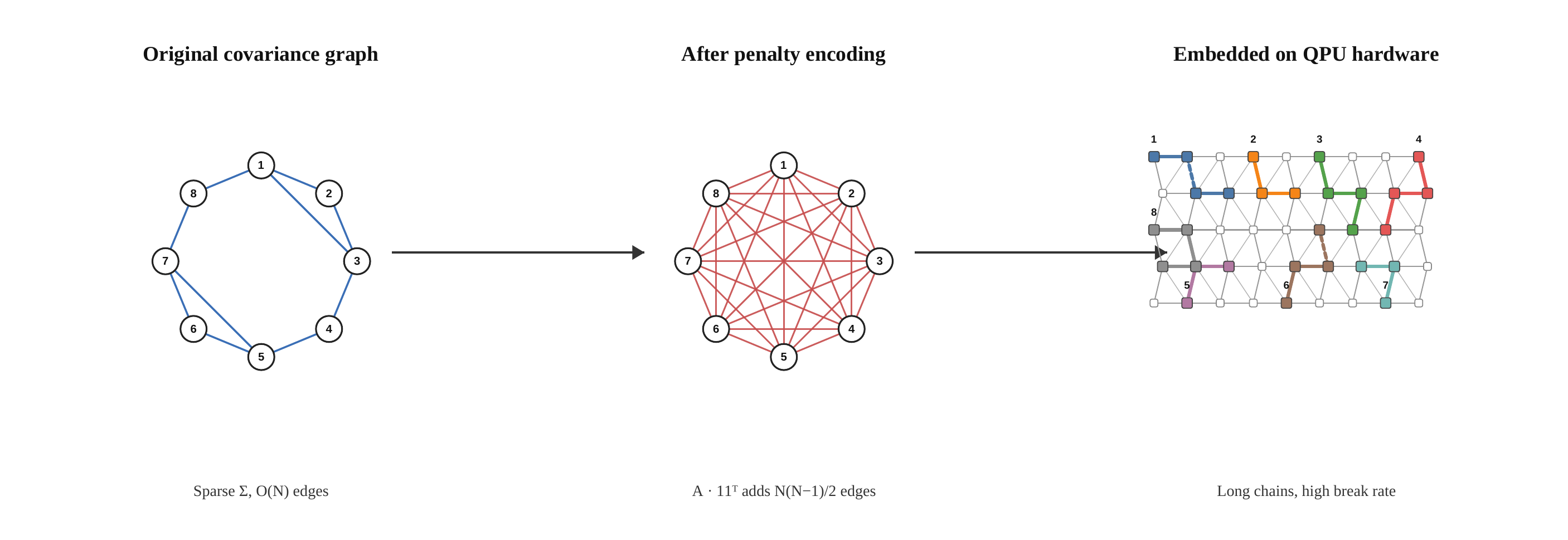}
    \caption{Penalty-dilution mechanism. Left: the original covariance
    graph may be sparse (few edges). Center: adding the cardinality penalty
    $A\bone\bone^\top$ makes the QUBO fully connected regardless of the
    original structure. Right: embedding the complete graph on
    bounded-degree hardware requires long chains that are prone to
    breaking during annealing.}
    \label{fig:penalty_dilution}
\end{figure}

Expanding the penalty and collecting terms, the problem takes the standard
QUBO form $f_Q(\bx) = \bx^\top Q\, \bx + c$, where
\begin{equation}
    Q = -\mathrm{diag}(\bmu) + \lambda \Sigma + A\, \bone\bone^\top - 2AK\, I
    \label{eq:qubo_matrix}
\end{equation}
and $c = AK^2$ is a constant offset. The key observation, and the source of
the difficulties we report in this paper, is that the penalty term
$A\, \bone\bone^\top$ contributes $A$ to \emph{every} off-diagonal entry of
$Q$, making the interaction graph complete regardless of the structure of
$\Sigma$. Even if the financial interactions are sparse (as they are in the
betting case below), the QUBO is guaranteed to be dense because of the
encoding. For equities, the covariance matrix $\Sigma$ is itself empirically
dense (all industries correlate), so the QUBO would be dense even without
the penalty. For betting, $\Sigma$ is block-diagonal and the penalty is the
sole source of density. This is a property of the penalty encoding, not of
portfolio optimization itself. The financial problem has sparse or
structured interactions; the QUBO surrogate does not.
Figure~\ref{fig:penalty_dilution} illustrates the mechanism: the penalty
term transforms any sparse interaction graph into a complete graph, which
then requires long embedding chains on bounded-degree hardware topologies.

\begin{figure}[t]
    \centering
    \includegraphics[width=\textwidth]{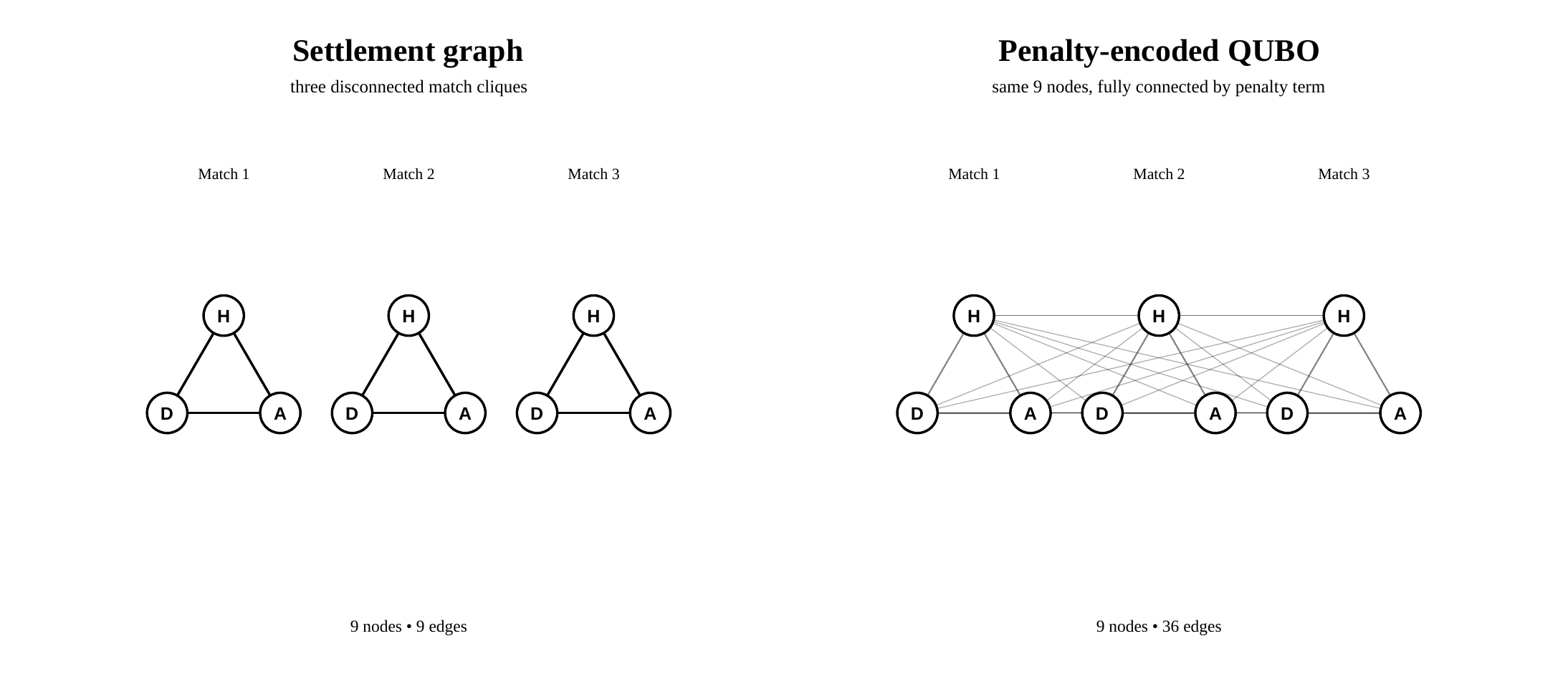}
    \caption{Settlement graph versus penalty-encoded QUBO for a 3-match
    betting slate ($N = 9$). Left: the settlement graph has 9 edges (three
    disconnected 3-cliques). Right: the penalty-encoded QUBO has 36 edges
    (complete graph $K_9$) because the cardinality penalty
    $A\bone\bone^\top$ connects every pair of nodes.}
    \label{fig:settlement}
\end{figure}

\paragraph{Case studies.}
We instantiate Equation~\eqref{eq:portfolio} on two public-data case
studies. For the equity case, $\bmu$ is the rolling mean daily return
estimated over a 252-trading-day window and $\Sigma$ is the sample
covariance matrix over the same window; we rebalance monthly. From the
Kenneth French 49 industry daily portfolios we construct subsets of size
$N \in \{12, 16, 20, 24, 32, 40, 49\}$ by ranking the 49 industries by
absolute mean return and selecting the top $N$, with the full $N=49$
universe used without subsetting at the largest scale. For the betting
case, for simultaneous pre-match bets with decimal odds $d_i$ and
de-vigged consensus probabilities
$p_i$~\citep{uhrin2021}, the expected return and covariance are
\begin{equation}
    \mu_i = d_i\, p_i - 1, \qquad
    \Sigma_{ij} = \begin{cases}
        d_i^2\, p_i(1 - p_i) & \text{if } i = j, \\
        -d_i\, p_i\, d_j\, p_j & \text{if } i \neq j \text{ and same match}, \\
        0 & \text{otherwise}.
    \end{cases}
    \label{eq:betting_cov}
\end{equation}
This gives $\Sigma$ a natural block-diagonal structure: within each match
the three mutually exclusive 1X2 outcomes form a clique, while cross-match
covariance is zero under the standard independence assumption.
Figure~\ref{fig:settlement} makes the contrast concrete: the settlement
graph has only $3M$ edges (one triangle per match of $M$ matches), while
the penalty-encoded QUBO has $\binom{3M}{2}$ edges.

\subsection{Topology-aware sparsification}\label{sec:formulation-sparsify}

A natural response to a dense logical graph on sparse hardware is to
sparsify the QUBO matrix itself: replace $Q$ with $\tilde{Q}$ obtained by
zeroing some off-diagonal entries, accept the constraint-violation cost,
and recover feasibility through classical post-processing. We study four
sparsification families.
\emph{Threshold} retains entries $|Q_{ij}| \geq \tau$ for a fixed cutoff
$\tau > 0$.
\emph{Top-$k$} retains, for each node $i$, the $k$ off-diagonal entries of
largest magnitude, symmetrized.
\emph{Domain-prior mask} retains only entries permitted by a boolean
template that encodes domain knowledge --- for equities a $k$-nearest-neighbor
correlation graph, for betting the exact settlement graph.
\emph{Domain-prior with residual edges} starts from the domain mask and
adds the top $r$ off-template entries ranked by absolute correlation.
Threshold and top-$k$ are generic graph-compression baselines; the
domain-prior methods carry the financial structure of the problem.
Diagonal entries are preserved by every sparsifier. Detailed sweeps over
$\tau$, $k$, and $r$ are reported in the supplement
(Online Resource A); the main text reports the four families as summary
points. A formal perturbation analysis showing that the spectral and
max-entry bounds on $\|Q - \tilde{Q}\|$ are too loose to predict
optimizer preservation at the relevant scales is given in
Online Resource G; in this paper we therefore rely on empirical
post-processed regret rather than perturbation bounds.

\subsection{The penalty--sparsification tension}\label{sec:tension}

A key structural finding of this work follows immediately from
Equation~\eqref{eq:qubo_matrix}. The exact-$K$ penalty adds $A$ to every
off-diagonal entry of $Q$. When any sparsification method zeroes
off-diagonal entries, it removes penalty weight proportional to the number
of eliminated edges:
\begin{equation}
    \sum_{(i,j)\, \in\, \text{removed}} A \;=\; A \cdot |\text{removed edges}|.
\end{equation}
This weakens the cardinality constraint. In our experiments the effect is
severe: sparsified QUBOs produce raw QPU samples with $\sum_i x_i \gg K$,
typically near all-ones vectors, because the penalty for selecting too many
assets has been partially erased. The tension is not specific to any
particular sparsification method or problem domain. It arises whenever a
penalty-encoded constraint produces dense off-diagonal contributions that
dominate the objective structure. We observe identical behavior in both
equity covariance portfolios and betting settlement portfolios.

\paragraph{Coefficient convention.} We use the symmetric $\bx^\top Q\, \bx$
convention throughout the derivation above, so $Q$ is a symmetric matrix and
off-diagonal pairs are accounted for through $Q_{ij}$ and $Q_{ji}$
symmetrically. When constructing a binary quadratic model for the D-Wave
Ocean toolchain, off-diagonal couplers are exchanged into the upper-triangular
convention, so the effective $x_i x_j$ coefficient seen by the sampler is
$Q_{ij} + Q_{ji} = 2 Q_{ij}$ for $i < j$. All ``logical-edge'' counts
reported in this paper refer to non-zero unordered pairs $\{i,j\}$, not to
the sum of $Q_{ij}$ and $Q_{ji}$ entries; chain-strength and chain-break
quantities are also reported per unordered pair as returned by the Ocean
sampler. This convention is consistent across all chain-strength sweeps and
all reported live-QPU runs.

\section{Methods}\label{sec:methods}

\subsection{Instances and data}\label{sec:methods-data}

\paragraph{Equities.}
We use the Kenneth French 49-industry daily portfolios~\citep{french_data}
and construct rolling instances with 252-day estimation windows and monthly
rebalancing. Subsets at $N \in \{12, 16, 20, 24\}$ are formed by ranking
the 49 industries by absolute mean return and selecting the top $N$; for
scaling experiments we additionally test $N \in \{32, 40, 49\}$ with
$K = 12$ and $\lambda = 1.0$, with the $N = 49$ instance using the full
FF49 universe without subsetting. Risk-aversion values are
$\lambda \in \{0.5, 1.0, 2.0\}$ and the penalty weight is $A = 4.0$
throughout. The domain-prior graph is built from $k$-nearest-neighbor
correlations on the estimation-window covariance.

\paragraph{Betting.}
We use pre-match football 1X2 odds from
\texttt{football-data.co.uk}~\citep{footballdata} for five European
leagues (English Premier League, La Liga, Serie A, Bundesliga, Ligue~1)
across seasons 2020/21 through 2024/25, totalling 8{,}981 matches.
Matchday slates are constructed with 3-day windows, with slate sizes
from 3 to 16 matches ($N = 9$ to $N = 48$ selections) and cardinalities
$K \in \{3, 5, 8, 10\}$. The domain-prior graph is the settlement graph:
disjoint 3-cliques, one per match.

The full offline experimental grid produced 918 preservation rows, 4{,}212
embedding rows, 1{,}620 domain-prior comparison rows, and 405 validation
rows. Live-QPU sampling is reported in Section~\ref{sec:results} on the
subset of instances at the frontier scales identified by the offline
grid.

\subsection{Solver, sampling, post-processing, and metrics}\label{sec:methods-protocol}

\paragraph{Hardware.}
Live experiments use D-Wave Advantage\_system4.1 (Pegasus topology) and
Advantage2\_system1.13 (Zephyr topology). Offline embedding benchmarks use
ideal Pegasus ($m=16$, 5{,}760 nodes) and ideal Zephyr ($m=12$, 5{,}400
nodes) graphs generated by D-Wave NetworkX~\citep{pelofske2025}. Live
solvers expose smaller active graphs due to manufacturing yield, and we use
the live working graphs for QPU sampling. Across the full experimental
range ($N \leq 49$ for equities, $N \leq 48$ for betting), every tested
instance embeds successfully on both topologies, so the relevant comparison
is embedding quality (overhead and chain length), not embeddability.
Table~\ref{tab:qpu_hygiene} consolidates the live-solver properties, the
sampling protocol, and the D-Wave Ocean toolchain defaults used in this
work.

\begin{table}[ht]
\centering
\caption{QPU-hygiene parameters. The three explicit sampler keyword
arguments (\texttt{num\_reads}, \texttt{chain\_strength},
\texttt{annealing\_time}) are set in the project source code; every other
parameter inherits its D-Wave Ocean or solver default and is reported
verbatim here for the editorial record.}
\label{tab:qpu_hygiene}
\begin{tabular}{ll}
\toprule
Parameter & Value \\
\midrule
\multicolumn{2}{l}{\emph{Hardware (Pegasus)}}\\
~~ Solver & Advantage\_system4.1 \\
~~ graph\_id & 01d07086e1 \\
~~ Active qubits / couplers & 5{,}627 / 40{,}279 \\
~~ Ideal-graph nodes ($m=16$) & 5{,}760 \\
\midrule
\multicolumn{2}{l}{\emph{Hardware (Zephyr)}}\\
~~ Solver & Advantage2\_system1.13 \\
~~ graph\_id & 01e1ea5685 \\
~~ Active qubits / couplers & 4{,}579 / 41{,}549 \\
~~ Ideal-graph nodes ($m=12$) & 5{,}400 \\
\midrule
\multicolumn{2}{l}{\emph{Explicit sampling protocol}}\\
~~ Reads per submission & 1{,}000 \\
~~ Anneal schedule & linear forward, $t_a = 20$~\textmu s \\
~~ Chain strengths swept & $\{0.5, 1.0, 2.0\}$ \\
~~ Offline embedding & minorminer \\
~~ Live embedding & fixed-embedding reuse via FixedEmbeddingComposite \\
\midrule
\multicolumn{2}{l}{\emph{Ocean / solver defaults (not overridden)}}\\
~~ \texttt{auto\_scale} & \texttt{True} \\
~~ \texttt{num\_spin\_reversal\_transforms} & 0 (no gauge averaging) \\
~~ \texttt{chain\_break\_method}$^{\dagger}$ & majority vote \\
~~ \texttt{answer\_mode} & histogram \\
~~ \texttt{flux\_drift\_compensation} & \texttt{True} \\
~~ \texttt{reduce\_intersample\_correlation} & \texttt{False} \\
~~ \texttt{programming\_thermalization} & solver default \\
~~ \texttt{readout\_thermalization} & solver default \\
\bottomrule
\end{tabular}

\medskip
{\footnotesize $^{\dagger}$ \texttt{chain\_break\_method} applies only to
embedded logical-QUBO sampling via FixedEmbeddingComposite; physical-QUBO
calls sample the physical graph directly and use no chain unembedding.}
\end{table}

\paragraph{Post-processing.}
Greedy feasibility projection is applied to every raw QPU sample. If
$\sum_i x_i > K$, the projector iteratively flips the selected variable
with the smallest marginal contribution to the objective from 1 to 0; if
$\sum_i x_i < K$, it iteratively flips the unselected variable with the
largest marginal contribution from 0 to 1; the procedure repeats until
$\sum_i x_i = K$. This deterministic step guarantees exact-$K$
feasibility on every post-processed sample, and is identical across
sparsified and penalty-free QUBOs so that solution-quality comparisons
between formulations are not confounded by differing post-processors.

\paragraph{Metrics.}
\emph{Objective regret} is reported relative to the best available offline
reference. For the general grid, this reference is exact brute-force
enumeration at $N \leq 16$ and greedy construction with 128 random restarts
for $N > 16$. For the frontier instances highlighted in the main results
(notably $N = 24$, $K = 8$), we additionally perform exact enumeration over
all $\binom{N}{K}$ feasible subsets (tractable up to
$\binom{24}{8} = 735{,}471$), so that the regret values reported in
Section~\ref{sec:results} are exact and not heuristic approximations.
Support Jaccard overlap and exact-$K$ feasibility rate are also reported.
Embedding metrics are success rate, physical qubits, qubit overhead ratio,
and mean and maximum chain length. QPU metrics are best returned energy,
feasible sample rate, chain-break fraction (mean fraction of broken chains
per sample, averaged over reads), and sample diversity. For equities the
primary financial metric is the realized daily Sharpe
ratio~\citep{sharpe1966} over the out-of-sample evaluation window with
95\% bootstrap confidence intervals (1{,}000 draws); we additionally
compute the Probabilistic Sharpe Ratio and minimum track record length
following \citet{bailey2012,bailey2014}, but the out-of-sample windows
($T \approx 21$ trading days) are shorter than the MinTRL in most cases,
so we do not base our main conclusions on PSR-based inference. For betting
the main-text economic metric is realized ROI; Brier and log-loss scores
are computed and retained in the supplement but, following
\citet{wunderlich2020}, we treat ROI as a returns-based diagnostic rather
than as a standalone proxy for predictive skill (see Section~\ref{sec:discussion}).

\section{Results}\label{sec:results}

\begin{figure}[t]
    \centering
    \includegraphics[width=\textwidth]{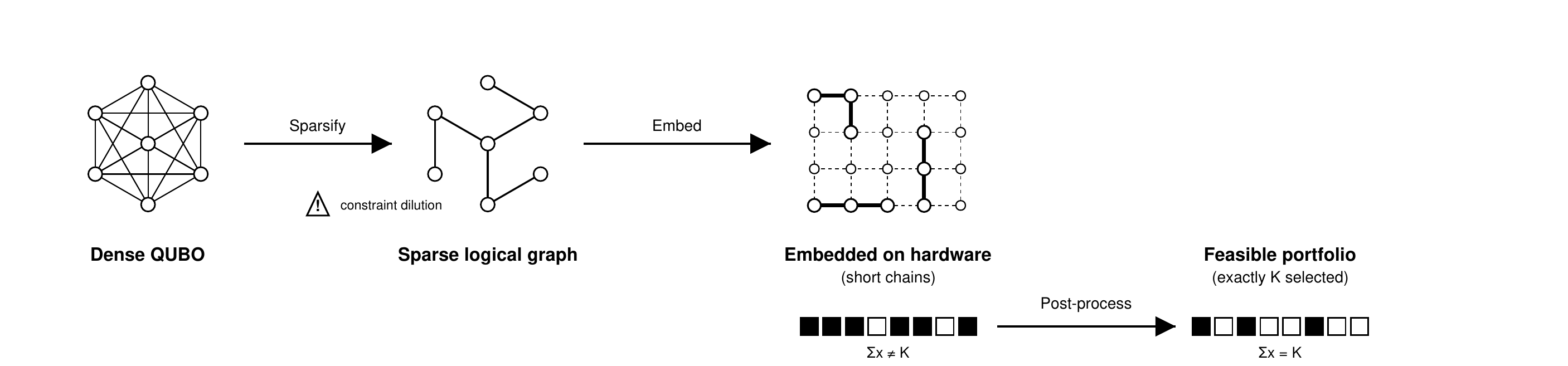}
    \caption{Three-stage direct-QPU pipeline for penalty-encoded portfolio
    QUBOs: sparsification improves embeddability, but induces constraint
    dilution, so feasibility-aware post-processing is required to recover
    exact-$K$ portfolios.}
    \label{fig:pipeline}
\end{figure}

We organize the results around the three-step diagnostic narrative of the
introduction: penalty-encoded direct-QPU sampling fails
(Section~\ref{sec:results-penalty-fails}); the natural fix of
topology-aware sparsification removes chain breaks but produces a pipeline
dominated by the classical projector
(Section~\ref{sec:results-sparsify}); removing the penalty entirely and
enforcing cardinality classically gives a pipeline whose chain breaks are
near zero and whose post-processed solutions are competitive with greedy
classical references at every tested scale
(Section~\ref{sec:results-penalty-free}).
Figure~\ref{fig:pipeline} sketches the three-stage penalized pipeline that
motivates Sections~\ref{sec:results-penalty-fails}
and~\ref{sec:results-sparsify}; the penalty-free pipeline of
Section~\ref{sec:results-penalty-free} replaces the first two stages with a
single objective-only sampling step followed by the same classical
projector.

\subsection{Penalty-encoded QUBOs fail on direct QPU}\label{sec:results-penalty-fails}

\begin{figure}[t]
    \centering
    \includegraphics[width=\textwidth]{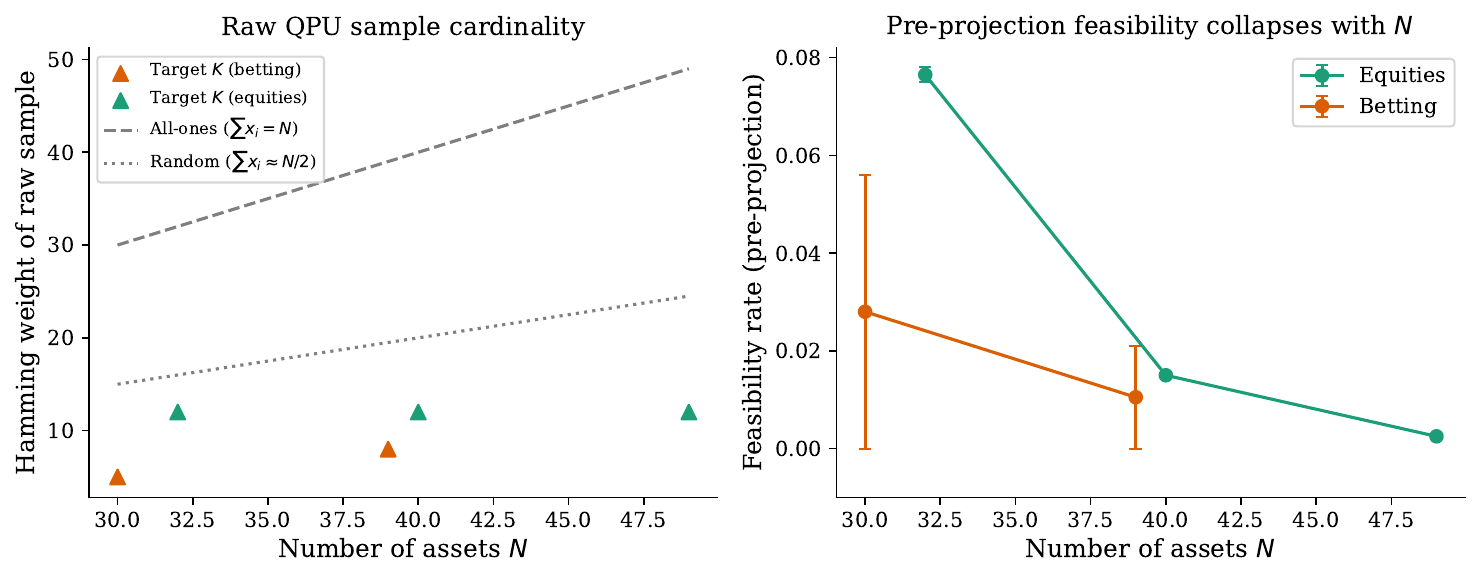}
    \caption{Raw QPU sample cardinality collapse under penalty encoding.
    Left: target cardinality $K$ relative to the all-ones ($N$) and random
    ($N/2$) references. Right: pre-projection feasibility rate collapses
    toward zero with increasing $N$, confirming that the QPU does not
    enforce the cardinality constraint at these chain-break rates.}
    \label{fig:cardinality_collapse}
\end{figure}

\begin{figure}[t]
    \centering
    \includegraphics[width=\textwidth]{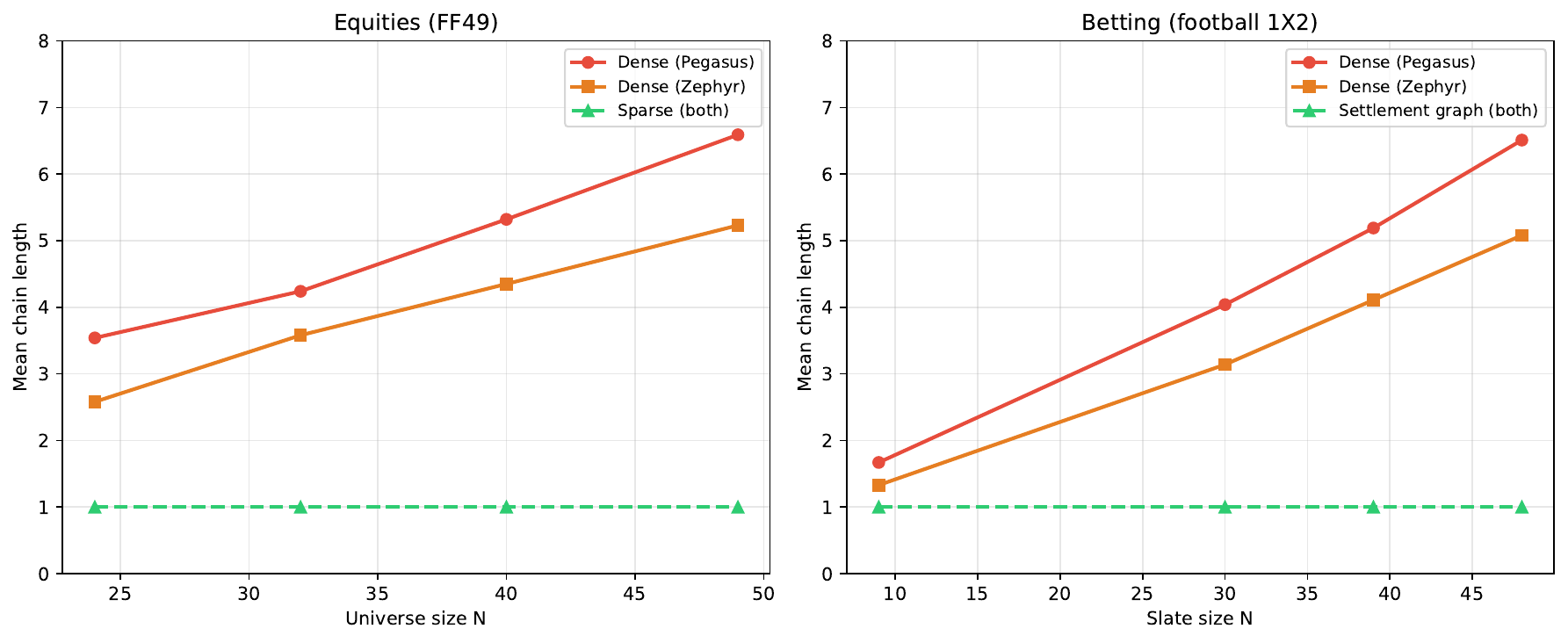}
    \caption{Mean chain length versus problem size for dense and
    best-sparse (top-$k$, $k = 1$ for equities; settlement graph for
    betting) QUBOs on both topologies. Dense chain lengths grow with $N$;
    the plotted sparse variants maintain unit chains at all scales.}
    \label{fig:scaling}
\end{figure}

Dense portfolio QUBOs produce chain-break fractions that grow steadily with
problem size. Figure~\ref{fig:cardinality_collapse} illustrates the core
diagnostic: raw QPU samples from penalty-encoded QUBOs fail to satisfy the
cardinality constraint, with the feasibility rate collapsing to zero as
$N$ grows. Table~\ref{tab:scaling} and Figure~\ref{fig:scaling} report the
full scaling trend across both case studies. At $N = 24$ ($K = 8$,
$\lambda = 2.0$), the dense equity instance required 85 physical qubits on
Pegasus (mean chain length 3.54) and 62 on Zephyr (mean chain length
2.58), with chain-break fractions of approximately 83\%. At $N = 49$ (the
full FF49 universe, $K = 12$), mean chain lengths grew to 6.59 on Pegasus
and 5.23 on Zephyr, with chain-break fractions of 88\% and 92\%
respectively. The feasible sample rate was 0\% at every scale tested: zero
out of thousands of reads returned exactly $K$ selected assets. Dense
betting QUBOs show the same pattern: chain breaks rise from 45\% at
$N = 9$ to 84\% at $N = 48$. Zephyr consistently embeds instances with
lower qubit overhead and shorter chains than Pegasus for the same logical
graph (full detail in Online Resource~C); the bottleneck at the scales we
tested is embedding quality, not embeddability.

\begin{table}[ht]
\centering
\caption{Scaling of embedding quality and live-QPU behavior with problem
size. Chain lengths are from ideal-topology embeddings; chain-break
fractions and post-processed regret (best of Pegasus and Zephyr) are from
live-QPU runs on Advantage\_system4.1 and Advantage2\_system1.13. Regret
is measured against the best greedy reference. Rows marked $^{\ast}$ have
ideal-embedding data only (no live-QPU run).}
\label{tab:scaling}
\begin{tabular}{llrrrrrrr}
\toprule
Case & $N$ & Sparsifier & Edges & \multicolumn{2}{c}{Mean chain} & \multicolumn{2}{c}{Chain break} & Regret \\
     &     &            &       & Peg. & Zep. & Peg. & Zep. & (post-proc.) \\
\midrule
\multicolumn{9}{l}{\textit{Equities (FF49, $K = 12$, $\lambda = 1.0$)}} \\
     & 32 & Dense &  496 & 4.24 & 3.58 & 87.5\% & 90.6\% & 0.32\% \\
     & 40 & Dense &  780 & 5.32 & 4.35 & 85.5\% & 81.8\% & 0.24\% \\
     & 49 & Dense & 1176 & 6.59 & 5.23 & 88.1\% & 91.8\% & 0.32\% \\
     & 49 & Top-$k$ ($k\!=\!1$) &  24 & 1.00 & 1.00 & 0.0\% & 0.0\% & 0.63\% \\
     & 49 & Domain-prior$^{\ast}$ &  82 & 1.17 & 1.09 & --- & --- & --- \\
\midrule
\multicolumn{9}{l}{\textit{Betting (football 1X2, $\lambda = 0.5$)}} \\
     & 30 & Dense &  435 & 4.04 & 3.14 & 70.7\% & 77.9\% & 22.1\% \\
     & 30 & Settlement &   30 & 1.00 & 1.00 & 0.0\% & 0.0\% & \textbf{0.0\%} \\
     & 39 & Dense &  741 & 5.19 & 4.11 & 84.6\% & 80.5\% & 34.0\% \\
     & 39 & Settlement &   39 & 1.00 & 1.00 & 0.0\% & 0.0\% & \textbf{0.0\%} \\
     & 48 & Dense & 1128 & 6.51 & 5.08 & 83.9\% & 83.9\% & 35.8\% \\
     & 48 & Settlement &   48 & 1.00 & 1.00 & 0.0\% & 0.0\% & \textbf{0.0\%} \\
\bottomrule
\end{tabular}
\end{table}

\subsection{Sparsify-and-project: chains fixed, constraint diluted}\label{sec:results-sparsify}

Sparsified versions of the same $N = 24$ frontier instance show
dramatically improved embedding (Table~\ref{tab:sparse_embedding}): all
four sparsification families reduce chain-break fractions to below the
per-sample resolution floor of $10^{-3}$ and bring physical-qubit counts
back to the same order as the logical instance. This pattern holds at
larger scales: the top-$k$ ($k = 1$) sparsifier at $N = 49$ achieves unit
chains and zero chain breaks on both solvers
(Table~\ref{tab:scaling}). Sparsification therefore solves the embedding
problem at every scale we tested.

\begin{table}[ht]
\centering
\caption{Embedding and chain behavior for the $N = 24$ equity frontier
instance on Advantage\_system4.1 (Pegasus). All chain-break fractions are
reported as the mean fraction of broken chains per sample, averaged over
$N_{\text{reads}}=1000$. The bound ``$<10^{-3}$'' denotes below the
per-sample detection floor of $1/N_{\text{reads}}$.}
\label{tab:sparse_embedding}
\begin{tabular}{lrrrrr}
\toprule
Sparsifier & Edges & Phys.\ qubits & Mean chain & Chain break & Feasible \\
\midrule
Dense                      & 276 & 85 & 3.54 & 83.3\%      & 0\% \\
Threshold                  &  69 & 35 & 1.46 & $<10^{-3}$  & 0\% \\
Top-$k$                    &  12 & 24 & 1.00 & $<10^{-3}$  & 0\% \\
Domain-prior               &  39 & 26 & 1.08 & $<10^{-3}$  & 0\% \\
Domain-prior + residuals   &  43 & 29 & 1.21 & $<10^{-3}$  & 0\% \\
\bottomrule
\end{tabular}
\end{table}

What sparsification does not do is restore feasibility. The feasible
sample rate remained 0\% for all methods at all scales. Inspection of the
raw samples revealed near-all-ones vectors ($\sum_i x_i \approx N$),
indicating that the exact-$K$ penalty was effectively neutralized by
sparsification, as analyzed in Section~\ref{sec:tension}. The betting case
study makes the embedding fix especially clean: settlement-graph
sparsification reduces chain breaks to exactly 0\% across all slate sizes
($N = 9$ to $N = 48$), all chain-strength settings, and both solvers,
because each match contributes an independent 3-clique with constant
maximum degree, so the settlement graph remains uniformly sparse with
linear edge growth ($3M$ edges for $M$ matches), in contrast to equity
covariance which densifies quadratically with the universe size. Yet the
feasible sample rate stays at 0\% in all betting cases, confirming that
the penalty--sparsification tension is structural and not specific to
covariance density.

\begin{figure}[t]
    \centering
    \includegraphics[width=0.75\textwidth]{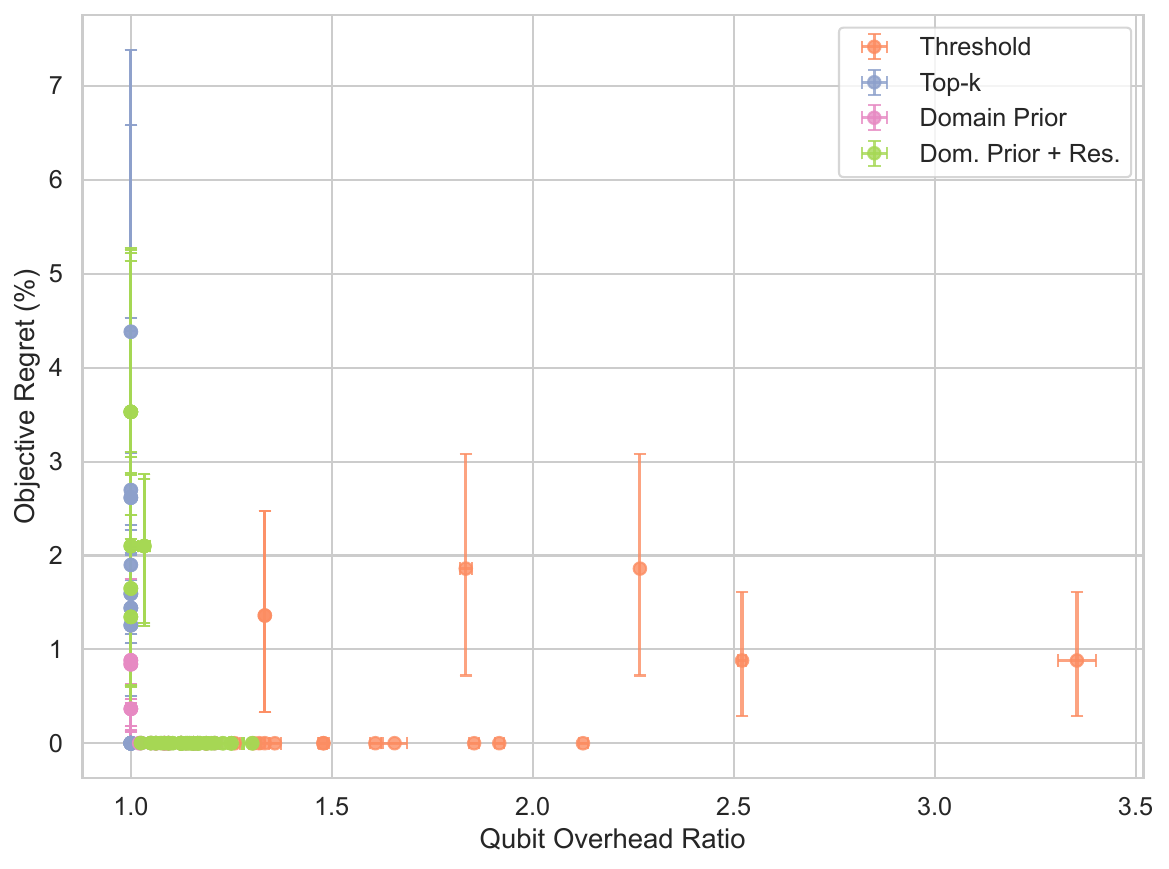}
    \caption{Objective regret versus qubit-overhead ratio on the offline
    sparsification grid. Lower-left is better. Domain-prior methods
    achieve low regret at low overhead, while threshold and top-$k$ trade
    more aggressively. The offline Pareto ranking is preserved by the
    embedding, but the live-QPU outcome is dominated by the classical
    projection step (see Figure~\ref{fig:ablation}).}
    \label{fig:regret_overhead}
\end{figure}

The quality of the post-processed solutions is the more interesting
question.
Figure~\ref{fig:regret_overhead} reports the offline regret--overhead
Pareto frontier for the sparsification families, distilled from the full
918-row preservation grid summarized in Online Resource~B. Domain-prior
sparsification on betting instances achieves near-zero regret because the
settlement graph captures the covariance block structure exactly. On
equities, the $k$-nearest-neighbor correlation prior preserves more
objective quality than threshold or top-$k$ at matched edge budgets. So
sparsifiers can be ranked offline, but offline ranking is not what
determines the live-QPU outcome.

\begin{figure}[t]
    \centering
    \includegraphics[width=0.85\textwidth]{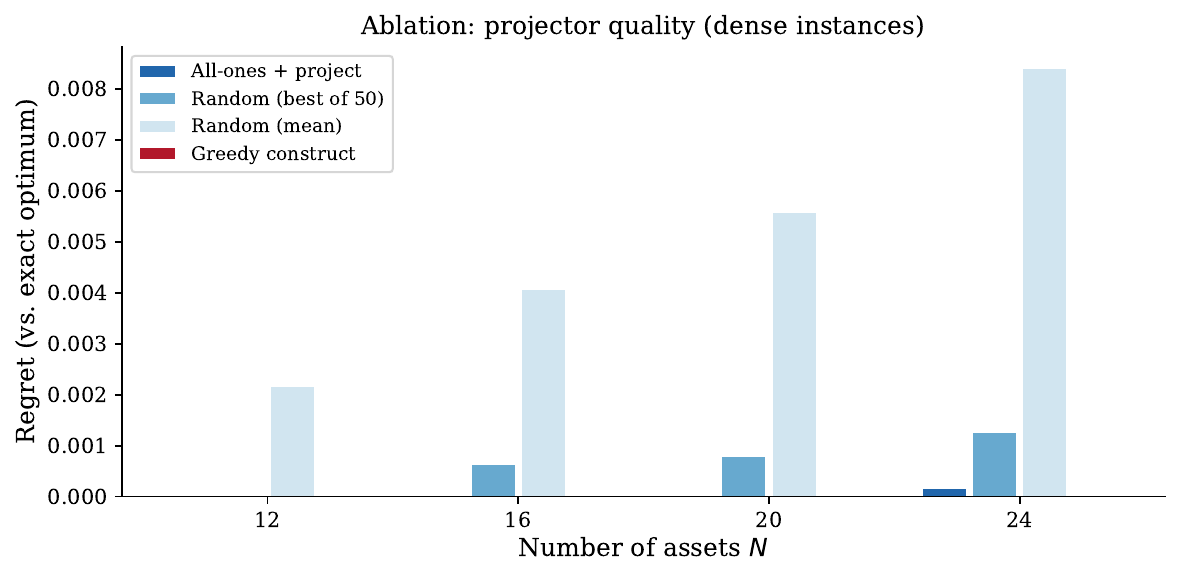}
    \caption{QPU vs.\ projector ablation. Random projection (mean)
    degrades with $N$, while all-ones projection and greedy construction
    achieve near-zero regret on penalty-encoded dense QUBOs, indicating
    that the classical projector contributes the bulk of the pipeline's
    output quality at the scales we tested.}
    \label{fig:ablation}
\end{figure}

The projector-dominance pattern is most stark on the betting case study
because the settlement-graph QUBO decomposes into independent 3-cliques
that a deterministic backward-elimination projector can solve to zero
regret from any reasonable starting vector. Table~\ref{tab:betting_ablation}
confirms this directly: on the settlement-graph QUBO, all-ones projection
matches the QPU-plus-projection result at zero regret with perfect support
overlap at all three sizes; on the dense-betting QUBO, all-ones projection
suffers regret of 39.4\% at $N=30$ rising to negative regret of $-62.2\%$
at $N=48$ (i.e., it finds a different feasible portfolio with even lower
energy than greedy, which is not a proven optimum at that size). The
zero-regret result on settlement betting is therefore a property of the
problem's decomposable structure and the projector, not of quantum
sampling.

\begin{table}[ht]
\centering
\caption{Betting ablation: post-processed regret and support Jaccard
versus the greedy reference. On the settlement-graph QUBO, all-ones
projection matches the greedy reference exactly at every scale, showing
that the zero-regret result is a property of the problem structure and
the projector, not of the QPU. The dense-betting rows establish the
counter-evidence: when the structure is removed, the projector's
behavior changes substantially.}
\label{tab:betting_ablation}
\begin{tabular}{lllrr}
\toprule
$N$ & QUBO & Method & Regret & Jaccard \\
\midrule
30 & Settlement & QPU + proj.\ (Pegasus)  & 0.0\%               & 1.000 \\
30 & Settlement & All-ones + proj.        & 0.0\%               & 1.000 \\
30 & Dense      & All-ones + proj.        & 39.4\%              & 0.250 \\
\midrule
39 & Settlement & QPU + proj.\ (Pegasus)  & 0.0\%               & 1.000 \\
39 & Settlement & All-ones + proj.        & 0.0\%               & 1.000 \\
39 & Dense      & All-ones + proj.        & 16.4\%              & 0.000 \\
\midrule
48 & Settlement & QPU + proj.\ (Pegasus)  & 0.0\%               & 1.000 \\
48 & Settlement & All-ones + proj.        & 0.0\%               & 1.000 \\
48 & Dense      & All-ones + proj.        & $-62.2\%^{\dagger}$ & 0.000 \\
\bottomrule
\end{tabular}

\medskip
{\footnotesize $^{\dagger}$ Negative regret indicates the all-ones
projection found a lower-energy solution than the greedy reference, which
is not a proven global optimum at this size.}
\end{table}

The same pattern holds on equities at $N = 24$, $K = 8$. On the dense
QUBO, greedy construction finds the exact optimum (zero regret); all-ones
projection achieves 0.017\% regret; the QPU plus post-processing achieves
0.514\% regret, the worst of all methods, because the 83\% chain-break
rate produces severely degraded raw samples that even post-processing
cannot fully repair. On the threshold-sparsified QUBO, the QPU plus
post-processing achieves 0.050\% regret while greedy construction
achieves 0.097\% --- but all-ones projection on the \emph{dense} QUBO
still achieves 0.017\%, better than both, because all-ones projection is
effectively backward elimination from the full universe and is therefore
a well-known greedy heuristic for cardinality-constrained quadratic
problems. The per-sparsifier equity rows are detailed in Online
Resource~B, S-Tab~1.

To be explicit about what we are and are not claiming for this pipeline:
at the scales we tested ($N \leq 49$), the QPU does not outperform simple
classical baselines on the sparsify-and-project pipeline of this section.
For betting, the zero-regret result is a property of the problem's
decomposable structure and the greedy projector, not of quantum sampling.
For equities, the QPU outperforms greedy construction on
threshold-sparsified QUBOs at $N = 24$, but simpler baselines (all-ones
projection on the dense QUBO) are competitive. The contribution of this
paper is not a performance claim for the QPU. It is a structural
diagnosis: the penalty encoding is what makes the QUBO dense,
sparsification is what makes direct-QPU access possible, and
post-processing is what recovers feasibility; but the QPU itself is not
yet the bottleneck or the differentiator at these scales. The next
subsection shows what happens when the penalty is removed instead of
sparsified.

\subsection{Penalty-free pipeline on live hardware}\label{sec:results-penalty-free}

The diagnosis of Section~\ref{sec:tension} and the ablation of
Section~\ref{sec:results-sparsify} motivate a direct test: if the penalty
is the source of density and constraint dilution, what happens if we
simply remove it? We build the objective-only QUBO
\begin{equation}
    Q_{\mathrm{obj}} = -\mathrm{diag}(\bmu) + \lambda\, \Sigma,
    \label{eq:qobj}
\end{equation}
submit it to the QPU without any penalty or sparsification, and enforce
the cardinality constraint through the same greedy feasibility projector
used in Section~\ref{sec:results-sparsify}. The key difference is that
post-processing is no longer repairing a dilution artifact; it is
enforcing a separate constraint on top of a correctly sampled objective
landscape.

We tested this pipeline on live Pegasus and Zephyr solvers at the same
scales used in the scaling analysis of
Section~\ref{sec:results-penalty-fails}. For each instance, we ran
1{,}000 reads at chain strengths $\{0.5, 1.0, 2.0\}$ and post-processed
every sample.

\paragraph{Betting ($\boldsymbol{\lambda = 0.5}$).}
The penalty-free betting QUBO is block-diagonal because same-match 1X2
outcomes form 3-cliques and cross-match covariance is zero under the
independence assumption. At $N \in \{30, 39, 48\}$, the number of
off-diagonal edges drops from $\binom{N}{2} \in \{435, 741, 1128\}$
(penalized) to exactly $N \in \{30, 39, 48\}$ (penalty-free). Ideal
Pegasus and Zephyr embeddings give unit chains at every scale, and live
chain-break fractions are essentially zero ($<10^{-4}$) across all
solvers and chain strengths. Post-processed regret relative to the greedy
classical reference is 0.0 at $N = 30$ (exact hit on all six runs),
$-0.429$ at $N = 39$, and $-0.944$ at $N = 48$. Negative regret indicates
that the QPU finds lower-energy feasible portfolios than greedy
construction; the greedy reference is not a proven global optimum, so
these gains should be read as improvements over a specific classical
heuristic, not as proofs of quantum advantage. Critically, the all-ones
projection baseline no longer produces zero regret on betting in the
penalty-free pipeline (0.394 at $N = 30$, 0.164 at $N = 39$): the
projector is no longer doing the entire job.

\paragraph{Equities ($\boldsymbol{\lambda = 1.0}$, $\boldsymbol{K = 12}$).}
The equity covariance matrix is empirically dense, so $Q_{\mathrm{obj}}$
remains a complete graph with $\binom{N}{2}$ off-diagonal edges. Removing
the penalty does not reduce the edge count, but it does reduce the dynamic
range of the couplings: the largest off-diagonal magnitude drops from
$A \approx 4.0$ (dominated by the penalty) to $\sim 10^{-3}$ (set by
$\lambda \Sigma$). This has a dramatic effect on embedding quality. Mean
chain lengths still grow with $N$ (4.1--6.2 on Pegasus, 3.3--4.9 on
Zephyr), but live chain-break fractions collapse from 83--92\% (penalized
equities) to at most 0.04\% on any solver and chain strength.
Post-processed regret relative to greedy construction is at most 0.03\%
($N = 49$), with several configurations achieving exact hits against
brute-force optima at $N = 24$ or matching the greedy reference at
$N = 40$.

\begin{table}[ht]
\centering
\caption{Penalty-free pipeline results. For each instance, we report the
number of off-diagonal edges in $Q_{\mathrm{obj}}$ versus the penalized
QUBO, the best live mean chain-break fraction across both solvers and all
chain strengths, and the best post-processed regret relative to the
greedy classical reference (brute-force exact reference at $N = 24$).}
\label{tab:penalty_free}
\begin{tabular}{llrrrr}
\toprule
Case & $(N, K)$ & $|E_{\mathrm{obj}}|$ & $|E_{\mathrm{penalized}}|$ & Chain break (min) & Best pp regret \\
\midrule
Equities & $(24, 8)$ & 276 & 276 & $0.0000$ & $0.000$ (exact) \\
Equities & $(32, 12)$ & 496 & 496 & $0.0000$ & $0.000116$ \\
Equities & $(40, 12)$ & 780 & 780 & $0.0000$ & $0.000000$ \\
Equities & $(49, 12)$ & 1176 & 1176 & $0.0000$ & $0.000270$ \\
\midrule
Betting & $(30, 5)$ & 30 & 435 & $0.0000$ & $0.000$ (matches greedy) \\
Betting & $(39, 8)$ & 39 & 741 & $0.0000$ & $-0.429$ (lower energy than greedy) \\
Betting & $(48, 10)$ & 48 & 1128 & $0.0000$ & $-0.944$ (lower energy than greedy) \\
\bottomrule
\end{tabular}
\end{table}

\begin{figure}[t]
    \centering
    \includegraphics[width=\textwidth]{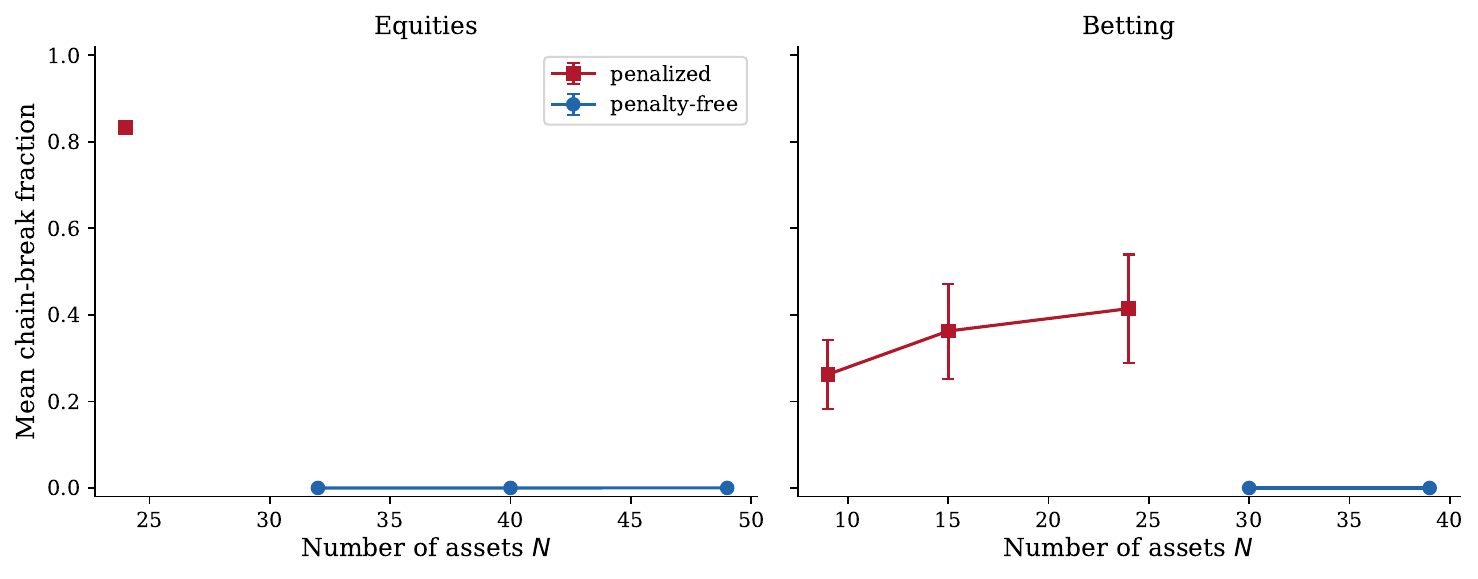}
    \caption{Head-to-head comparison of chain-break fractions: penalized
    pipeline (red) versus penalty-free pipeline (green) across equity and
    betting scaling instances on live D-Wave hardware. The same hardware,
    the same datasets, the same chain-strength sweeps, and the same
    post-processing routine produce chain-break fractions more than three
    orders of magnitude smaller when the penalty is removed.}
    \label{fig:penalized_vs_free}
\end{figure}

The contrast with the penalized pipeline is stark, as
Figure~\ref{fig:penalized_vs_free} makes visually explicit: the same
hardware, the same datasets, the same chain-strength sweeps, and the same
post-processing routine produce chain-break fractions that are more than
three orders of magnitude smaller when the penalty is removed, and
post-processed regret that is competitive with (equities) or better
than (betting) the greedy classical reference. We interpret this as
direct evidence that the penalty encoding, not the sparse hardware
topology, is the binding constraint at the scales we tested. The
mechanism by which removing the penalty restores chain integrity is not
that the topology becomes more permissive (chain lengths still grow with
$N$ in the penalty-free equity case), but that the coupling dynamic range
collapses by roughly three orders of magnitude --- from $A \approx 4.0$ in
the penalized objective to $\sim 10^{-3}$ in $\lambda\, \Sigma$ ---
pulling the largest objective coupling well below the chain-coupling
magnitude and stabilizing the chains. Topology will become the next
binding constraint at larger scales; at $N \leq 49$ it is not.

\paragraph{Sanity checks.}
Two robustness checks confirm the result is not an artifact of a
particular solver state or evaluation window. First, representative
instances were rerun on both live solvers and produced consistent
chain-break behavior and embedding statistics across Advantage\_system4.1
and Advantage2\_system1.13. On 2026-04-10 D-Wave renamed
\texttt{Advantage2\_system1.13} to \texttt{Advantage2\_system1} and
removed qubit 4374 (and its couplers) from the working graph; scanning
every saved embedding record produced for this paper confirms that qubit
4374 was never selected by the embedder for any chain on any reported
instance, so the rename does not affect any result reported here. Second,
on the same 21-instance out-of-sample validation grid the penalty-free
pipeline's hardware-and-objective win does not translate into short-window
financial dominance: the penalty-free live-QPU pipeline posts realized
Sharpe 0.035 versus 0.209 for the best sparse threshold equity method and
realized ROI 0.090 versus 0.200 for the best sparse domain-prior betting
method (full S-Tabs in Online Resource~E). The out-of-sample windows
($T \approx 21$ trading days) are shorter than the minimum track record
length computed via the PSR framework in most configurations, so these
differences are directionally informative rather than statistically
distinguishable. The hardware-and-objective result is what is being
claimed in this section; the short-window financial reading is not.

\section{Discussion}\label{sec:discussion}

Our experiments traced two distinct pipelines for direct-QPU portfolio
optimization. The first is the natural response to the observed
chain-break failures: sparsify the logical graph, embed the sparse graph,
and post-process raw samples to enforce feasibility. This pipeline does
produce feasible portfolios on live hardware, and at the $N \leq 49$
scales we tested it is competitive with classical heuristics, but the
ablation in Section~\ref{sec:results-sparsify} showed that its output is
dominated by the classical projection step --- on betting with
settlement-graph priors, an all-ones starting vector plus greedy
projection yields identical results without any QPU involvement. The
second pipeline, introduced in Section~\ref{sec:results-penalty-free}, is
simpler and more direct: skip the penalty encoding entirely, sample the
objective-only QUBO $Q_{\mathrm{obj}}$ on hardware, and handle the
cardinality constraint through the same classical projector. This
eliminates the penalty--sparsification tension by removing the penalty
upstream, not by working around it downstream. The empirical consequences
are concrete: live chain-break fractions drop from 71\%--92\% (penalized,
across both case studies) to at most 0.04\% across all tested scales;
the all-ones projection baseline is no longer trivially optimal for
betting; and the QPU makes a visible contribution to the final portfolio
quality (betting at $N = 39$ and $N = 48$ yields lower-energy feasible
portfolios than the greedy heuristic, an energy comparison not a proof
of optimality, while equities stay within 0.03\% regret through
$N = 49$).

The mechanism behind the penalty--sparsification tension is elementary,
and it bears restating in one place. The exact-$K$ penalty
$A(\bone^\top\bx - K)^2$ expands to a dense rank-one
contribution $A\bone\bone^\top$ that adds $A$ to every off-diagonal entry
of the QUBO matrix, regardless of whether the corresponding pair of
assets has any financial interaction. Even if the financial covariance
$\Sigma$ is sparse (as in the betting case), the total QUBO $Q$ is
guaranteed to be dense because of the encoding. ``Penalty-preserving''
sparsification --- zeroing only the objective entries while keeping the
penalty intact --- does not help, because $A\bone\bone^\top$ is itself
dense, so the resulting QUBO still has $\binom{N}{2}$ non-zero
off-diagonal entries. We swept chain strengths $\{0.5, 1.0, 2.0\}$
independently of the penalty weight $A$ because the focus of this paper
is structural diagnosis, not parameter optimization; a coupled
penalty-aware chain-strength sweep is a natural follow-on, as are
alternative structural encodings such as one-hot $K$-subset
representations that enforce cardinality through the variable structure
rather than through a quadratic penalty.

The deeper observation is that what worked in
Section~\ref{sec:results-penalty-free} was \emph{separating two roles}
that are conflated in the standard penalty-encoded QUBO. The QPU is used
only to sample the risk--return objective landscape; the exact-$K$
investment constraint is enforced deterministically afterward. This
separation matters because the final object handed to a practitioner is
an investable $K$-asset portfolio, not an infeasible low-energy binary
vector. We conjecture that the observation generalizes beyond portfolio
optimization: any combinatorial optimization problem currently handled by
submitting a penalty-encoded QUBO to a sparse-topology QPU should benefit
from separating the objective (sampled on hardware) from constraints
(enforced classically), provided the constraint admits a fast projection.
Verifying that conjecture on other problem classes is a direction for
future work. We did not apply spin-reversal-transform gauge averaging
during sampling (Table~\ref{tab:qpu_hygiene}); the failure mode under
analysis is the completeness of the logical interaction graph, not the
distribution of intrinsic-control-error biases, so SRT averaging is not
load-bearing for the qualitative finding.

We are explicit about the scale of these results. Our equity experiments
span $N = 12$ to $N = 49$ (the full FF49 universe), and our betting
experiments span $N = 9$ to $N = 48$. At these scales, both Pegasus and
Zephyr embed all instances successfully, dense or sparse; the bottleneck
is embedding quality, not embeddability. The topology argument becomes
practically compelling only when $N$ approaches the embedding capacity
of the hardware: extrapolating from our observed scaling (mean chain
length growing roughly as $0.13N$ on Pegasus), chain lengths would
exceed 8 at $N \approx 60$ and 12 at $N \approx 90$, making reliable
sampling increasingly difficult. These are back-of-the-envelope
extrapolations from the observed data, not established thresholds, and
the precise limits depend on the specific QUBO coefficients, the
embedding algorithm, and the working graph. What we can say with
confidence is that the scaling trend is monotonically worsening for
dense penalty-encoded QUBOs while remaining flat for the penalty-free
formulation. For the equity case, exact enumeration over
$\binom{49}{12} \approx 2.6 \times 10^{11}$ subsets is expensive but
feasible with branch-and-bound, and greedy heuristics find high-quality
solutions in milliseconds; the QPU does not offer a computational
advantage at these scales. Block-diagonal financial structures such as
event-settlement graphs preserve the topology-friendly property at any
scale, but the corresponding classical problem is also typically
decomposable, so the same caution applies.

A note on financial evaluation. We report realized daily Sharpe ratio
with bootstrap confidence intervals as the primary equity metric, and
realized ROI as the primary betting metric. With evaluation windows of
approximately 21 trading days, the minimum track record length computed
via the PSR framework of \citet{bailey2012,bailey2014} substantially
exceeds our window length in most cases, which means that observed
differences in Sharpe ratios between methods are directionally
informative rather than statistically significant performance claims. In
the apples-to-apples 21-instance validation grid summarized in
Section~\ref{sec:results-penalty-free}, the penalty-free live-QPU
pipeline is the hardware-and-objective winner but is not the best
realized financial row: Online Resource~E reports realized Sharpe
0.035 for penalty-free versus 0.209 for the best sparse threshold equity
method, and realized ROI 0.090 versus 0.200 for the best sparse
domain-prior betting method. We do not view this as contradicting the
main result, because the central claim concerns structural viability on
direct annealing hardware, objective fidelity, and constraint handling
rather than short-window financial dominance. Following
\citet{wunderlich2020}, we caution against interpreting realized betting
ROI as evidence of predictive skill: short-window positive ROI can arise
without a superior forecasting model, the so-called profitability
paradox. The Brier and log-loss diagnostics computed for our betting
selections are reported in Online Resource~E for completeness.

\section{Conclusion}\label{sec:conclusion}

In the present work, we asked why direct portfolio optimization fails on
current quantum annealers and whether the failure can be fixed at
currently accessible scales.

Our experiments showed that the standard penalty-encoded exact-$K$ QUBO
contains a dense rank-one term $A\bone\bone^\top$ that makes the logical
interaction graph complete regardless of the underlying financial
structure, producing chain-break fractions of 83\% at $N = 24$ rising to
88--92\% at $N = 49$ and zero feasible raw samples on D-Wave Pegasus and
Zephyr hardware. They showed that the most natural remedy ---
topology-aware sparsification followed by classical feasibility projection
--- does reduce chain breaks to essentially zero, but also dilutes the
cardinality constraint, and that on structurally favorable cases (betting
with settlement-graph priors) the post-processed portfolio quality is
explained by the classical projector rather than by the QPU samples. And
they showed that simply dropping the penalty and sampling the
objective-only QUBO
$Q_{\mathrm{obj}} = -\mathrm{diag}(\bmu) + \lambda \Sigma$ on hardware
reduces live chain-break fractions to at most 0.04\% across the full
experimental range ($N \leq 49$ for equities, $N \leq 48$ for betting),
matches the greedy reference on equities to within 0.03\% regret, and
returns lower-energy feasible portfolios than greedy on betting at
$N \in \{39, 48\}$ (an energy comparison, not a proof of optimality).
The all-ones projection baseline that explained the sparsify-and-project
betting result no longer suffices in the penalty-free pipeline.

We do not claim quantum advantage. The greedy reference is not a proven
global optimum at larger scales, and the QPU's wins on betting should be
read as improvements over a specific classical heuristic. What we do
claim is more modest and more useful: within the tested exact-$K$
portfolio formulations and at currently accessible scales ($N \leq 49$),
direct quantum-annealer portfolio optimization became practically viable
only after removing the penalty encoding from the QUBO and handling the
cardinality constraint classically through feasibility projection. The
common practice of submitting penalty-encoded QUBOs to the QPU is the
reason the direct pipeline fails in the literature we reviewed; a simple
modification fixes it for the class of problems and scales we tested.

Several extensions are natural. Sparse event-settlement workflows
(prediction-market basket sizing, event-driven hedging) share the
block-diagonal structure of the betting case and should benefit directly
from the same separation of sampling and constraint enforcement.
Multi-period rebalancing with turnover and transaction-cost constraints
would test whether the same penalty-free separation remains useful
across time, and ESG-constrained or mandate-constrained portfolios would
test whether multiple practitioner constraints can be handled through
structured projection rather than dense penalty terms. More broadly, the
same approach is worth exploring on any direct-QPU combinatorial
optimization problem where a binary constraint is currently enforced
through a quadratic penalty that dominates the off-diagonal structure of
the QUBO.

\backmatter

\bmhead{Acknowledgements}
The author thanks the D-Wave team for providing access to quantum computing
resources through the Leap cloud platform and for their continuous
technical support. Live QPU experiments used \texttt{Advantage\_system4.1}
(Pegasus) and \texttt{Advantage2\_system1.13} (Zephyr).

\section*{Declarations}

\textbf{Funding.} This research received no external funding. QPU access
was obtained through a D-Wave Leap subscription.

\textbf{Competing interests.} The author declares no competing interests.

\textbf{Author contributions.} L.L.\ is the sole author and conceived the
study, designed the experiments, implemented the code, ran the
experiments, analyzed the results, and wrote the manuscript.

\textbf{Data availability.} The Fama--French 49-industry daily portfolios
used in this study are publicly available from the Kenneth R.\ French Data
Library; pre-match football 1X2 odds are publicly available from
\texttt{football-data.co.uk}. Synthetic mean-variance instances were
generated programmatically with fixed random seeds reported in the
manuscript and are reproducible from the source code, released under the
MIT License and publicly available at
\url{https://github.com/LuisLozanoM/penalty-free-portfolio}.

\textbf{AI-assistance disclosure.} AI-based coding assistants and
editorial tools were used for code scaffolding and language support; all
experimental design, verification, analysis, and final manuscript
decisions were performed by the author.

\bibliography{references}

\end{document}